\documentclass[showpacs,preprintnumbers,prd,nofootinbib,floats,amssymb,floatfix]{revtex4}
\usepackage{amsmath}
\usepackage{amsfonts}
\usepackage{graphicx}
\usepackage{hyperref}
\usepackage{amsmath}
\usepackage{amstext}
 
\begin{document}

\title {Dirac and Faddeev-Jackiw quantization  of a five-dimensional  St{\"{u}}eckelberg theory  with a  compact dimension}
 \author{ Alberto Escalante}  \email{aescalan@ifuap.buap.mx}
\author{ Moises Zarate}  \email{mzarate@ifuap.buap.mx}
\affiliation{ Instituto de F{\'i}sica Luis Rivera Terrazas, Benem\'erita Universidad Aut\'onoma de Puebla, (IFUAP). Apartado
 postal J-48 72570 Puebla. Pue., M\'exico,}

\begin{abstract}
%We study the St{\"{u}}ckelberg's Hamiltonians in the presence of an extra dimension 
%compactified on a circle of radius $R$ ($M^{4}\times S^{1}$ spacetime). 
A detailed Hamiltonian analysis for a five-dimensional  St{\"{u}}eckelberg  theory   with a  compact dimension is performed. First, we develop a pure Dirac's analysis of the theory,  we show that after  performing  the compactification,   the theory is reduced  to  four-dimensional  St{\"{u}}eckelberg theory  plus a tower of Kaluza-Klein modes.  We  develop a  complete analysis of the constraints, we fix  the gauge and we show that there are present   pseudo-Goldstone bosons. Then we quantize the theory by   constructing  the Dirac brackets. As complementary work, we perform the Faddeev-Jackiw quantization for the theory under study,  and we calculate the generalized Faddeev-Jackiw brackets,   we show that both the Faddeev-Jackiw  and  Dirac's  brackets are the same. Finally we discuss some  remarks and prospects. 
\end{abstract}

\date{\today}
\preprint{}
\maketitle
\section{INTRODUCTION}
\vspace{1em} \
 It is well know that four dimensional Proca's  theory is not a gauge theory;  the theory describes a massive vector field and the physical degrees of freedom are three, this is, the addition of a mass term to Maxwell theory breaks the gauge invariance of the theory and adds one physical degree of freedom to electromagnetic degrees of freedom \cite{1a, 1,2}.  However, in spite of  Proca    is not a gauge theory and it was believed by several people   that only massless vector theories are  gauge invariant,   St{\"{u}}eckelberg  introduced to Proca's theory a scalar field converting the   theory  to  be massive but preserving  gauge invariance \cite{3b, 3}. The St{\"{u}}eckelberg's mechanism consists  in the  introduction of new fields to reveal a symmetry of a gauge fixed theory.  Moreover, Pauli  showed that St{\"{u}}eckelberg's formulation of  a massive vector field satisfies  a restricted U(1) gauge invariance, similar to that one encountered in quantum electrodynamics \cite{4, 5}.  The studio of  St{\"{u}}eckelberg's Lagrangian becomes to be relevant in several contexts  of theoretical physics,  for instance, gauge bosons masses through St{\"{u}}eckelberg couplings are present in string theory and  supergravity \cite{6},  the mechanism turned out to be crucial in the covariant quantization of the spacetime supersymmetric string theory \cite{7},  and also  St{\"{u}}eckelberg  fields were  introduced   as an essential tool for the formulation of the antisymmetric partner to the graviton \cite{8}. Furthermore,  St{\"{u}}eckelberg's mechanism provides an alternative way  to  the Higgs mechanism. In fact, St{\"{u}}eckelberg's mechanism archive gauge symmetry braking without affecting  renormalizability \cite{9}. \\
By taking into account the ideas explained above, in this paper we perform the canonical analysis   and the Faddeev-Jackiw [FJ] quantization  for  a five-dimensional St{\"{u}}eckelberg  theory   with a compact   dimension.  Nowadays,   the study of models involving extra dimensions  have an   important activity  in order to explain and solve  some  fundamental  problems found in theoretical physics,  such as,  the problem of  mass hierarchy,   the explanation of dark energy,  dark matter and inflation  \cite{10}. Furthermore, extra dimensions become also  important  in theories of  unification trying  of incorporating   gravity and gauge interactions consistently, for instance, string theory and grand unification theories. Moreover,  there are phenomenological and theoretical motivations to quantize a gauge  theory in extra dimensions, for instance, if  there exist extra dimensions, then  their  effects  could be tested in the actual  LHC collider, and  in the International Linear Collider \cite{11}. In this manner,  the study of five-dimensional  St{\"{u}}eckelberg theory  becomes relevant in the context of extra dimensions, in particular, we study the effects of the extra dimension on the theory  when it is compactified on a $S^1/\mathbf{Z_2}$ orbifold. In this respect, let us show the relevance  that the higher Kaluza-Klein [KK] modes  of a vector field gain their masses  through a St{\"{u}}eckelberg's  mechanism by performing the  compactification  on a circle.  In fact,  in \cite{3} is reported the following action
\begin{equation}
L_5D= -\frac{1}{4} \mathcal{F}_{IJ}(x_I) \mathcal{F} ^{IJ}(x_I) - \frac{1}{2 \xi} \left( \partial_I \mathcal{A}^I (x_I) \right)^2, 
\label{1}
\end{equation}
where  $x_I(x_\mu, y )$, $x_\mu$ label the four-dimensional manifold and $y$ the fifth-spatial extra dimension, $\xi$ is as usual a gauge parameter, $\mathcal{A}_I= (A_\mu(x_I), \theta(x_I)) $ is the five-dimensional Abelian gauge field and $\theta(x_I) $ is the St{\"{u}}eckelberg scalar. Now, we perform   the expansion of the gauge field in harmonics, namely 
\begin{eqnarray}
A_\mu(x_I)&=& \sum_{n=0}^{\infty} A_\mu^{(n)}(x_\mu) \zeta_n(y), \nonumber \\
\theta(x_I)&=&\sum_{n=0}^{\infty} \theta^{(n)}_\mu(x_\mu) \eta_n(y),\nonumber 
\end{eqnarray}
where $\zeta$ and $\eta$ are harmonic functions on the interval $(0, 2\pi R)$. Thus, by considering that expansion and integrating over the fifth dimension we obtain   the following four-dimensional  Lagrangian
\begin{eqnarray}
L_{4D}= \sum_{n=0}^{\infty} \left[ -\frac{1}{4} \mathcal{F}_{\mu \nu}^{(n)}\mathcal{F}^{\mu \nu}_{(n)} -\frac{n^2}{2} \left( \frac{1}{R} A_\mu^{(n)} + n\partial_\mu \theta^{(n)} \right)^2  -\frac{1}{2\xi}    \left( \left(\partial_\mu A^{\mu (n)} \right)^2 + \frac{2n}{R} \partial_\mu A^{\mu (n)} \theta^{(n)}  +\left(\frac{1}{R} \right)^2 (\theta^{(n)})^2 \right)   \right], 
\label{es1}
\end{eqnarray} 
by choosing $\xi=1$ we find that the $\theta^{(n)}$ field decouple from the vector fields.  Thus, we obtain one massless vector field and an infinite tower of massive vector fields, obtaining their mass by means the St{\"{u}}eckelberg mechanism, and there is no Higgs mechanism involved in the generation of mass. \\
Hence, our study will be focussed   in the  St{\"{u}}eckelberg theory  \cite{3b}, we will perform the compactification on a $S^1/\mathbf{Z_2}$ orbifold and   the theory under study   will involve     massive zero-modes,    a tower of massive vector fields and pseudo-Goldstone bosons.
 Our analysis  will be carry out  by  performing   a pure Dirac's method \cite{12, 13, 14, 15}; we develop  a full constraints program and we construct the Dirac brackets of the theory. On the other hand,  we also perform the [FJ] symplectic  formulation \cite{16};  we show that Dirac's and generalized [FJ] brackets are the same. We apply the [FJ] formulation because  it is a fundamental method for quantization; in fact,  the [FJ]  framework has been applied in several singular theories but it has  not been applied for studying  theories with compact  dimensions,  and    we show in this paper  that it  is an elegant framework  for annalyzying  theories  in this context. \\ 
The paper is organized as follows: In Sect. II,  we study  a five-dimensional   St{\"{u}}eckelberg theory, after performing  the compactification process on a  $S^1/\mathbf{Z_2}$ orbifold we obtain a  four-dimensional  effective Lagrangian. We perform the Hamiltonian analysis and we obtain the complete constraints of the theory. By using the constraints, we show that under an appropriate fixed  gauge,  the fields $A_5^{(n)}$  are identified as  pseudo-Goldston bosons, just like it  appears  in five-dimensional Maxwell  theory. In Sect.III, we
compute the Dirac brackets of the zero-modes. In Sect. IV, we calculate the Dirac brackets of the Kaluza-Klein [KK] modes . In Sect. V, we perform the [FJ] method; we construct the generalized [FJ] brackets and we show  that the obtained [FJ]  and Dirac's brackets coincide  for the zero-modes. In Sect. VI, we calculate the generalized [FJ] brackets for the KK-modes; we show the equivalence among [FJ] and Dirac's brackets.  In Sect. VII,  we  present some remarks and  prospects. 
 \newline
\newline
\section{Dirac's analysis  for  five-dimensional  St{\"{u}}eckelberg  theory with a compact dimension}
The action that we shall study in this section is given by the following  five-dimensional   St{\"{u}}eckelberg theory \cite{3b}
\begin{equation}
S \left[ A(x_{\mu},y),\theta(x_{\mu},y)\right]= \int d^{4}x \int_{0}^{2\pi R} dy \left\{-\frac{1}{4} F{^{M N}}F_{M N}+m^{2}\left(A_{M}+\partial_{M}\theta\right)\left(A^{M}+\partial^{M}\theta\right)\right\}, 
\label{ac1}
\end{equation}
here $\theta$ is the St{\"{u}}eckelberg scalar. It is important to remark, that the action is invariant under the gauge transformations 
 \begin{eqnarray}
 A_M(x, y) &\rightarrow &A_M(x, y) - \partial_M \epsilon(x,y), \nonumber \\
 \theta(x, y) &\rightarrow & \theta(x, y) + \epsilon(x,y), 
 \label{gaugetrans}
 \end{eqnarray}
 and the compactification will be performed in order  to do not damage that gauge symmetry. \\
The notation that we will use along the paper is the following; the capital latin indices $M, N$ run over $0,1,2,3,5$ here $5$ label the extra compact dimension and these indices  can be raised and lowered by the five-dimensional Minkowski metric $\eta_{M N}= (1,-1,-1,-1,-1)$; $y$ will represent the coordinate in the compact dimension and $\mu, \nu=0,1,2,3$ are spacetime indices, $x^\mu$  the coordinates that label the points for the four-dimensional manifold $M_4$; furthermore we will suppose that the compact dimension is a $S^1/\mathbf{Z_{2}}$ orbifold whose radius is $R$. In this respect, it is well knew that a simple compact one dimensional manifold is a circle $S^1$,  just as was developed the compactification above. However, if we  demand  an additional reflection symmetry $ Z_2$ with respect to the origin $y=0$, then we obtain an  orbifold $S^1 / Z_2$  which turns out to be important in the  study of higher dimensional physics. The point $y=0$ is a fixed point because it is $Z_2$ invariant, and  also  $-\pi R$ is a second fixed point of the orbifold, thus, we can observe that $S^1/Z_2$  reduces to $S^1$ to line segment with fixed endpoints at $y=0$ and $y=\pi R$.   Hence, the periodic boundary conditions on $S^1$  indicate that any dynamical field on $S^1/Z_2$ must be expanded in terms of functions with period  $0$ to $2\pi R$. Therefore, the compactification on $S_1/Z_2$ reflects certain  restrictions on the fields  and by taking account the gauge symmetry (\ref{gaugetrans}), we requires that the fields satisfy  the following  \cite{18}
\begin{eqnarray}
A_M(x,y)&=& A_M(x, y+2\pi R), \nonumber \\
A_\mu(x, y)&=& A_\mu (x, -y), \nonumber \\
A_5(x, y)&= &-A_5(x, -y),  \nonumber \\
\theta(x, y)&=& \theta(x, y+2\pi R), \nonumber\\ 
\theta(x, y)&=& \theta(x, -y).  
%\epsilon(x,y ) &=& (x, y+2\pi R), \nonumber \\
%\epsilon(x,y ) &=& (x, -y), 
\end{eqnarray}
It is important to comment that for $S^1$ we expanded  the fields in terms of complex exponentials, however,  in that expansion there are mixtures of even and odd functions and  this is not an appropriate  basis for $S^1/Z_2$. Nevertheless,  the dynamical variables defined on  $M_{4}\times S^1/\mathbf{Z_2}$ can be expanded in terms of the complete set of harmonics  \cite{18, 17, 19}
\begin{eqnarray*}
A_{5}(x,y)& = &\frac{1}{\sqrt{\pi R}}\sum_{n=1}^{\infty}A^{(n)}_{5}(x)\sin \left( \frac{n y}{R} \right),\nonumber\\
A_{\mu}(x,y)& = &\frac{1}{\sqrt{2\pi R}}A^{(0)}_{\mu}(x) + \frac{1}{\sqrt{\pi R}}\sum_{n=1}^{\infty}A^{(n)}_{\mu}(x)\cos \left(\frac{n y}{R} \right), \nonumber\\
\theta(x,y)=& = &\frac{1}{\sqrt{2\pi R}}\theta^{(0)}(x) + \frac{1}{\sqrt{\pi R}}\sum_{n=1}^{\infty}\theta^{(n)}(x)\cos \left(\frac{n y}{R} \right).
\end{eqnarray*}
For this theory, the dynamical variables for the zero mode are given by  $A^{(0)}_{i},A^{(0)}_{0},\theta^{(0)}$ and for the KK-modes   are $A^{(n)}_{5},A^{(n)}_{i},A^{(n)}_{0},\theta^{(n)}$ with   $i,j=1,2,3$. We shall  suppose that the number of KK-modes   is $k$, and we will  take the limit $k \rightarrow \infty$  at the end of the calculations, thus,  $n=1, 2, 3...k-1$.\\
Let us develop the Hamiltonian  analysis for the  action (\ref{ac1});  hence, we perform the $4+1$ decomposition and the compactification  process, thus,  the Lagrangian density takes the following form    
\begin{eqnarray}
{\mathcal{L}}&=&-\frac{1}{4}F^{(0)}_{\mu\nu}F^{\mu\nu}_{(0)}+m^{2}\left(A^{(0)}_{\mu}+\partial_{\mu}\theta^{(0)}\right)\left(A^{\mu}_{(0)}
+\partial^{\mu}\theta_{(0)}\right)+
\sum_{n=1}^{\infty}\Bigg[-\frac{1}{4}F^{(n)}_{\mu\nu}F^{\mu\nu}_{(n)}\nonumber\\
& &+m^{2}\left(A^{(n)}_{\mu}+\partial_{\mu}\theta^{(n)}\right)\left(A^{\mu}_{(n)}+\partial^{\mu}\theta_{(n)}\right)
+\frac{1}{2}\Big(\partial_{\mu}A^{(n)}_{5}+\frac{n}{R}A^{(n)}_{\mu}\Big)(\partial^{\mu}A^{5}_{(n)}+\frac{n}{R}A^{\mu}_{(n)}\Big) \nonumber\\
& &-m^{2}\Big(A^{(n)}_{5}-\frac{n}{R}\theta^{(n)}\Big)(A^{(n)}_{5}-\frac{n}{R}\theta_{(n)}\Big)\Bigg], 
\label{eqlaga}
\end{eqnarray}
where $F^{(0)}_{\mu \nu}$ and $F^{(n)}_{\mu \nu}$ are the field strength associated with the fields $A^{(0)}_{\mu}$ and  $A^{(m)}_{\mu}$ respectively. To proceed with the canonical analysis,  we define the momenta  $\left( \Pi^{M}_{(n)},P_{(n)}\right)$ conjugate to the fields $\left(A^{(n)}_{M},\theta^{(n)}\right)$  respectively in the usual way 
\begin{equation}
\Pi^{M}_{(n)}= \frac{\delta L}{\delta \dot{A}^{(n)}_{M} },\;\; P_{(n)}= \frac{\delta L}{\delta \dot{\theta}^{(n)} },
\label{eq47}
\end{equation}
hence 
\begin{eqnarray}
\Pi^{0}_{(0)}&=&0,\nonumber\\
\Pi^{i}_{(0)}&=&\partial_{0}A^{(0)}_{i}-\partial_{i}A^{(0)}_{0},\nonumber\\
P_{(0)}&=&2m^{2}\left(A^{(0)}_{0}+\partial_{0}\theta^{(0)}\right),\nonumber\\
\Pi^{0}_{(n)}&=&0,\nonumber \\
\Pi^{i}_{(n)}&=&\partial_{0}A^{(n)}_{i}-\partial_{i}A^{(n)}_{0},\nonumber \\
\Pi^{5}_{(n)}&=&\partial_{0}A^{(n)}_{5}+\frac{n}{R}A^{(n)}_{0},\nonumber \\
P_{(n)}&=&2m^{2}\left(A^{(n)}_{0}+\partial_{0}\theta^{(n)}\right).
\label{48c}
\end{eqnarray}
 It is straightforward  observe that the Hessian of the the action (\ref{eqlaga}) is singular, the rank of Hessian is $5k-6$ and  there is $k$ null vectors. Thus,   from the null vectors, we obtain the following $k$ primary constraints 
\begin{eqnarray}
\phi^{0}_{(0)} &:&\Pi^{0}_{(0)} \approx  0,\nonumber\\
\phi^{0}_{(n)} &:&\Pi^{0}_{(n)} \approx  0.
\label{eq61}
\end{eqnarray}  
The canonical Hamiltonian is obtained by the  Legendre transformation as
\begin{eqnarray}
H_{c}&=&\int d^{3}x\Bigg(\frac{1}{2}\Pi^{i}_{(0)}\Pi^{i}_{(0)}+\frac{1}{4m^{2}}P_{(0)}P_{(0)}+\frac{1}{4}F^{(0)}_{i j}F^{i j}_{(0)}-A^{(0)}_{0}\big(\partial_{i}\Pi^{i}_{(0)}+P_{(0)}\big)-m^{2}\left(A^{(0)}_{i}+\partial_{i}\theta^{(0)}\right)^{2}\nonumber\\
& &+\sum_{n=1}^{\infty}\Bigg[\frac{1}{2}\Pi^{i}_{(n)}\Pi^{i}_{(n)}+\frac{1}{4m^{2}}P_{(n)}P_{(n)}+\frac{1}{4}F^{(n)}_{i j}F^{i j}_{(n)}-A^{(n)}_{0}\Big(\partial_{i}\Pi^{i}_{(n)}+\frac{n}{R}\Pi^{5}_{(n)}+P_{(n)}\Big)
-m^{2}\left(A^{(n)}_{i}+\partial_{i}\theta^{(n)}\right)^{2}\nonumber\\
& &+\frac{1}{2}\Pi^{(n)}_{5}\Pi^{(n)}_{5}+\frac{1}{2}\big(\partial_{i}A^{(n)}_{5}+\frac{n}{R}A^{(n)}_{i}\big)^{2}
+m^{2}\left(A^{(n)}_{5}-\frac{n}{R}\theta^{(n)}\right)^{2}\Bigg]\Bigg),
\label{hami}
\end{eqnarray}
the addition of  primary constraints to the canonical Hamiltonian allows us to identify the primary   Hamiltonian 
\begin{eqnarray}
H_{P}&= &H_{c} + \int dx^3 \Big[\lambda^{(0)}_{(0)}\phi^{0}_{(0)}+\sum_{n=1}^{\infty}\lambda^{(n)}_{0}\phi^{0}_{(n)}\Big]. \nonumber\\
\label{eq81} 
\end{eqnarray}
The non-vanishing fundamental Poisson brackets are given by 
\begin{eqnarray}
\{A^{(n)}_{M}(x^0, x), \Pi^{N}_{(n)} (x^0, z) \} &=& \delta{^{N}}_{M} \delta^3(x-z), \nonumber \\ 
\{\theta^{(n)}(x^0, x), P_{(n)}(x^0, z) \} &=&\delta^3(x-z). 
\label{eq91}
\end{eqnarray}
Now, we need  identify  if the theory presents   secondary constraints; for this aim, we calculate consistency among the primary constraints, thus, we obtain the following secondary constraints 
\begin{eqnarray}
\dot{\phi}^{0}_{(0)}&= &\{\phi^{0}_{(0)}(x), {H}_{P} \} \approx 0 \quad \Rightarrow \quad \psi_{(0)}=\partial_{i}\Pi^{i}_{(0)}+P_{(0)}\approx 0,\nonumber\\
%   & & \nonumber\\
\dot{\phi}^{0}_{(n)}&= &\{\phi^{0}_{(n)}(x), {H}_{P} \} \approx 0 \quad \Rightarrow \quad \psi_{(n)}=\partial_{i}\Pi^{i}_{(n)}+\frac{n}{R}\Pi^{5}_{(n)}+P_{(n)}\approx 0.
\label{eq36}
\end{eqnarray}
On the other hand, from consistency of  secondary constraints,  does not emerge more constraints. In this way, with all the constraints at hand we need to identify which ones correspond to first  and second class. For this aim, we compute the Poisson brackets between the primary and  secondary constraints. We find that the  Poisson brackets between  primary and secondary constraints are computed as 
\begin{eqnarray}\label{eq13-1}
 \{\psi^{(0)}(y),\phi^{0}_{(0)}(x)\}& =&0,\nonumber \\
\{ \psi^{(n)}(y),\phi^{0}_{(m)}(x)\}& =&0,  
\end{eqnarray}
we observe that  the Poisson  brackets between  primary with  secondary constraints vanish, hence,  the constraints of the theory under study  are all first class constraints. In fact, there are 2   first class constraints for the zero-mode given by 
\begin{eqnarray}
\gamma^{0}_{(0)}&=&\Pi^{0}_{(0)} \approx 0, \nonumber \\
\gamma^{(0)} &=&\partial_{i}\Pi^{i}_{(0)}+P_{(0)} \approx 0, 
\label{eq41}
\end{eqnarray}
where the second constraint of (\ref{eq41}) is  identified as the Gauss constraint for the zero-mode of  conventional St{\"{u}}eckelberg   theory.  Furthermore, there are $ 2k-2$ first class constraints for the KK-modes  given by 
\begin{eqnarray}
\gamma^{0}_{(n)}&=&\Pi^{0}_{(n)} \approx 0,\nonumber \\
\gamma^{(n)}&=& \partial_{i}\Pi^{i}_{(n)}+\frac{n}{R}\Pi^{5}_{(n)}+P_{(n)}\approx 0,
\label{eq42}
\end{eqnarray}
where the second constraint of (\ref{eq42}) can be identified as the Gauss constraint for the excited  modes.\\
In this manner, we perform the counting of physical degrees of freedom as follows; there are $12k-2$ dynamical variables and $2k$ independent first class constraints, thus  
\begin{eqnarray*}
\text{Number of physical degrees of freedom}&=&\frac{1}{2}\left(12k-2 -2(2k)\right)\nonumber\\
&=& 4k-1, \nonumber
\end{eqnarray*} 
we observe that if  $k=1$, then we  obtain  $3$ physical degrees of freedom corresponding to the physical degrees of freedom for the St{\"{u}}eckelberg theory without a compact  dimension, these degrees of freedom are    associated with the  zero-mode as is expected \cite{3b, 3, 4}.\\
Moreover, by using the first class constraints obtained in  (\ref{eq41}), (\ref{eq42}) we find  the extended action 
\begin{eqnarray}
& &S_{E}\Big(Q_{K},P_{K},\lambda_{K}\Big)
=\int\Big[\dot{A}_{\nu}^{(0)}\Pi^{\nu}_{(0)}+\dot{\theta}^{(0)}P_{(0)}-\mathcal{H}^{(0)}-\beta_{(0)}\gamma^{(0)}
-\lambda^{(0)}_{0}\gamma^{0}_{(0)}
+\sum_{n=1}^{\infty} \Big\{ \dot{A}_{N}^{(n)}\Pi^{N}_{(n)}+\dot{\theta}^{(n)}\Pi_{(n)}\nonumber\\
& &-\mathcal{H}^{(n)}-\lambda^{(n)}_{0}\gamma^{0}_{(n)}- \beta_{(n)}\gamma^{(n)}\Big\} \Big]dx^3,
\label{actmax}
\end{eqnarray}
where   $Q_{K}$ y $P_{K}$ represent all the dynamical variables and their    canonical momenta respectively,    $\lambda_{K}$ stand for  all  Lagrange multipliers associated with the first class constraints. From the extended action, we identify the  extended Hamiltonian given by
\begin{eqnarray}
H_{ext}=H_{c}+\int\Big[\beta_{(0)}\gamma^{(0)}+\lambda^{0}_{(0)}\gamma^{(0)}_{0}+\sum_{n=1}^{\infty} \Big\{\lambda^{(n)}_{0}\gamma^{0}_{(n)}+\beta_{(n)}\gamma^{(n)}\Big\}\Big]dx^3.
\end{eqnarray}
Now,  we will calculate the gauge transformations on the phase space. For this important step, we use the first class constraints and we   define  the following gauge generator 
\begin{eqnarray}
G= \int_\Sigma \left[\varepsilon^{0}_{(n)} \gamma^{(n)}_{0}  +  \varepsilon_{(n)} \gamma^{(n)} +\varepsilon^{0}_{(0)}\gamma^{(0)}_{0} + \varepsilon_{(0)}\gamma^{(0)} \right]dx^3, 
\label{eq62}
\end{eqnarray}
thus, we find that the   gauge transformations on the phase  space  for the zero-mode given by 
\begin{eqnarray}
\delta A^{(0)}_{\mu} &=& -\partial_{\mu}\varepsilon^{(0)},  \nonumber \\
\delta \theta^{(0)} &=&\varepsilon^{(0)},  
\label{eq63}
\end{eqnarray}
and the gauge transformation for the KK-modes  takes the form
\begin{eqnarray}
\delta A^{(n)}_{\mu} &=&-\partial_{\mu}\varepsilon^{(n)},  \nonumber \\
\delta A^{(n)}_{5} &=&  \frac{n}{R}\varepsilon^{(n)},  \nonumber \\
\delta \theta^{(n)}&=&\varepsilon^{(n)}, 
\label{eq17v}
\end{eqnarray}
we can observe that this result is in agreement  with the transformations (\ref{gaugetrans}).  On the other hand,  from the gauge transformations  (\ref{eq17v}) we can  consider   the particular   gauge fixing defined by $\varepsilon^{(n)}=-\frac{R}{n}A_{5}^{(n)}$. By using this gauge,   the effective action given in (\ref{eqlaga}) is reduced  to
\begin{eqnarray}
{\mathcal{L}}&=&-\frac{1}{4}F^{(0)}_{\mu\nu}F^{\mu\nu}_{(0)}+m^{2}\left(A^{(0)}_{\mu}+\partial_{\mu}\theta^{(0)}\right)\left(A^{\mu}_{(0)}+\partial^{\mu}\theta_{(0)}\right)+
\sum_{n=1}^{\infty}\Bigg[-\frac{1}{4}F^{(n)}_{\mu\nu}F^{\mu\nu}_{(n)}+\big(m^{2}+\frac{n^{2}}{2R^{2}}\big)A^{(n)}_{\mu}A^{\mu}_{(n)}\nonumber\\
& &+2m^{2}A^{(n)}_{\mu}\partial^{\mu}\theta_{(n)}
+m^{2}\partial_{\mu}\theta^{(n)}\partial^{\mu}\theta_{(n)}-\frac{m^{2}n^{2}}{R^{2}}\theta^{(n)}\theta_{(n)}\Bigg],  
\label{eqlagal2}
\end{eqnarray}
where we are  able to observe that the KK-modes are massive   fields, and $A_{5}^{(n)}$ has been absorbed and  identified as   pseudo-Goldstone bosons, just like is present in free five-dimensional  Maxwell theory \cite{17,18}. Therefore, the five dimensional St{\"{u}}eckelberg theory  with a compact dimension,  is composed by  a four-dimensional  St{\"{u}}eckelberg theory associated with the zero-mode,  a tower of  KK-modes   $A_{\mu}^{(n)}$  of mass $m^{2}+\frac{n^{2}}{2R^{2}}$, and a tower of massive  KK-modes  $\theta^{(n)}$  of mass $\frac{m^{2}n^{2}}{R^{2}}$  plus interactive terms.\\
In the following sections, we will quantize the theory by constructing the Dirac brackets,  then we will perform the [FJ] quantization of systems with constraints,  and we shall prove that  Dirac's  and the generalized [FJ] brackets are the same. We will find that the advantage for  applying the [FJ]  is  that there are   less steps for arriving to the generalized brackets in comparison  with Dirac's method, all this will be explained   along the paper. 
\section{Dirac's bracket for the zero-modes}
In this section we will quantize the theory. By following with  Dirac's formalism,  the first class constraints  obtained for the zero-mode and the KK-modes  are given by 
\begin{eqnarray}
\gamma^{0}_{(0)}&=&\Pi^{0}_{(0)} \approx 0, \nonumber \\
\gamma^{(0)} &=&\partial_{i}\Pi^{i}_{(0)}+P_{(0)} \approx 0, \nonumber \\
\gamma^{0}_{(n)}&=&\Pi^{0}_{(n)} \approx 0,\nonumber \\
\gamma_{(n)}&=& \partial_{i}\Pi^{i}_{(n)}+\frac{n}{R}\Pi^{5}_{(n)}+P_{(n)}\approx 0,
\label{eq19a}
\end{eqnarray}
since  the zero-mode and the KK-modes are not coupled  in both  the Lagrangian and  the  constraints,  then we can   construct  the Dirac brackets independently  for each case, namely,   first  we will construct the Dirac brackets for  the  zero-mode,  then we will construct the brackets  for the KK-modes.\\
 In order to construct the Dirac brackets for the zero-mode, we work with  the following   fixed   gauge $ \partial^i A_i^{(0)} \approx 0$ and $A_{0}^{(0)}\approx 0$,  obtaining  the following set of constraints
\begin{eqnarray}
\chi_1^{(0)}&=& \partial^i A_i^{(0)} \approx 0, \nonumber\\ 
\chi_2^{(0)} &=&\partial_{i}\Pi^{i}_{(0)}+P_{(0)} \approx 0, \nonumber \\
\chi^{0}_{3(0)}&=&\Pi^{0}_{(0)} \approx 0, \nonumber \\
\chi^{0}_{4(0)}&=& A_{0}^{(0)}\approx 0,
\label{c22}
  \end{eqnarray}
  under  these gauge,  now  the constraints are  all  of second class. So, the $4\times 4$ matrix whose entries are formed by the Poisson brackets among the constraints (\ref{c22}),  namely $C_{\alpha \beta}$, is given by 
\begin{eqnarray}
\label{} C_{\alpha \beta}&=&
 \bordermatrix{
                 &  &  &  &         \cr
 & \{\chi_1^{(0)}(x), \chi_1^{(0)}(y) \}                & \{\chi_1^{(0)}(x), \chi_2^{(0)}(y) \}                 & \{\chi_1^{(0)}(x), \chi_3^{(0)}(y) \}                & \{\chi_1^{(0)}(x), \chi_4^{(0)}(y) \}                                       \cr
 & \{\chi_2^{(0)}(x), \chi_1^{(0)}(y) \}               & \{\chi_2^{(0)}(x), \chi_2^{(0)}(y) \}                 & \{\chi_2^{(0)}(x), \chi_3^{(0)}(y) \}            & \{\chi_2^{(0)}(x), \chi_4^{(0)}(y) \}                        \cr
 & \{\chi_3^{(0)}(x), \chi_1^{(0)}(y) \}                & \{\chi_3^{(0)}(x), \chi_2^{(0)}(y) \}               & \{\chi_3^{(0)}(x), \chi_3^{(0)}(y) \}                 & \{\chi_3^{(0)}(x), \chi_4^{(0)}(y) \}                         \cr
 & \{\chi_4^{(0)}(x), \chi_1^{(0)}(y) \}                 & \{\chi_4^{(0)}(x), \chi_2^{(0)}(y) \}                   & \{\chi_4^{(0)}(x), \chi_3^{(0)}(y) \}                & \{\chi_4^{(0)}(x), \chi_4^{(0)}(y) \}                             \cr}   \nonumber  \\ &=&\bordermatrix{
                 & \chi_1 ^{(0)}(y) & \chi_2^{(0)} (y)  & \chi_3 (y) ^{(0)}& \chi_4^{(0)} (y)           \cr
\chi_1 ^{(0)}(x) & 0                & \nabla^2                 & 0                & 0                                       \cr
\chi_2^{(0)} (x) & -\nabla^2                & 0                 & 0             & 0                         \cr
\chi_3 ^{(0)}(x) & 0                & 0              & 0                & -1                        \cr
\chi_4^{(0)}(x) & 0                & 0                 & 1                & 0                           \cr} \delta^3(x-y),
\end{eqnarray}
and  its inverse  takes the following form 
\begin{equation*}
\label{} C^{\alpha \beta}=
\bordermatrix{
                 &  &  &  &         \cr
 & 0                & \frac{1}{-\nabla^2}                 & 0                & 0                                       \cr
 & \frac{1}{\nabla^2}                & 0                 & 0             & 0                         \cr
 & 0                & 0              & 1                & 0                        \cr
 & 0                & 0                 & 0               & -1                           \cr} \delta^3(x-y).
\end{equation*}
In this manner, the Dirac brackets of two functionals $F$, $G$ defined on the phase space,  is expressed by
\[
\{F(x),G(z)\}_{D}\equiv\{F(x),G(z)\}-\int d^{2}ud^{2}w\{F(x),\xi_{\alpha}(u)\}C{^{\alpha\beta}}\{\xi_{\beta}(w),G(z)\},
\]
where $\{F(x),G(z)\}$ is the Poisson bracket  between two functionals $F,G$,  and $\xi_{\alpha}= (\xi_1, \xi_2, \xi_3, \xi_4)$ represent  the set of second class constraints. By using this fact, we obtain the following  Dirac's brackets for the zero-mode 
\begin{eqnarray}
\{A^{(0)}_i(x), \Pi^j _{(0)}(z)\}_{D} &=& \left( \delta^j{_{i}} - \frac{\partial ^j \partial_i}{ \nabla ^2} \right) \delta^3 (x-z), \nonumber \\ 
\{P_{(0)}(x), \theta^{(0)}(z)\}_{D} &=& - \delta^3 (x-z), \nonumber\\
\{\Pi^i_{(0)} (x), \theta^{(0)}(z)\}_{D} &=& \frac{\partial^i}{\nabla^2} \delta^3(x-z),
\label{eq21a}
\end{eqnarray}
these are the  Dirac brackets for a four-dimensional   St{\"{u}}eckelberg theory  \cite{19a}. 
\section{Dirac's brackets  for the KK-modes }
Now, we will calculate the Dirac brackets for the KK-modes. For this aim, we observe that the  gauge transformations  given in (\ref{eq17v}),  allow us to  work with the following  axial gauge  $A_5^{(n)} \approx 0$ and $\Pi_5 ^{(n)}+ \frac{n}{R}A_0^{(n)} \approx 0$.  Thus,   the  set of  constraints for the KK-modes  are  of second class  given by 
\begin{eqnarray}
\chi_1^{(n)}&=&A_5^{(n)} \approx 0,\nonumber \\
\chi_{2(n)}&=& \partial_{i}\Pi^{i}_{(n)}+\frac{n}{R}\Pi^{5}_{(n)}+P_{(n)}\approx 0, \nonumber \\
\chi_{3 (n)}&=&\Pi^0_{(n)}\approx 0, \nonumber \\
\chi_4^{ (n)}&=& \Pi_5^{(n)}+\frac{n}{R}A_0^{(n)} \approx 0.
\label{eq22a}
\end{eqnarray}
Hence, the $4\times 4$ matrix, namely $C^{(n)}_{\alpha \beta}$ , whose entries are given by the Poisson brackets among the constraints (\ref{eq22a}) is given by 
\begin{equation*}
\label{} C^{(n)}_{\alpha \beta}=
\bordermatrix{
                 & \chi_1^{(n)} (y) & \chi_2 ^{(n)}(y)  & \chi_3^{(n)} (y) & \chi_4 ^{(n)}(y)           \cr
\chi_1^{(n)} (x) & 0                & \frac{n}{R}                 & 0                & 1                                       \cr
\chi_2^{(n)} (x) & -\frac{n}{R}                & 0                 & 0             & 0                         \cr
\chi_3^{(n)} (x) & 0                & 0              & 0                & -\frac{n}{R}                        \cr
\chi_4^{(n)}(x) & -1                & 0                 & \frac{n}{R}                & 0                           \cr} \delta^3(x-z),
\end{equation*}
and  its inverse  takes the following form 
\begin{equation*}
\label{} \left(C^{(n)}_{\alpha \beta}\right)^{-1}=
\bordermatrix{
                 &  &   &  &            \cr
& 0                & -\frac{R}{n}               & 0                & 0                                       \cr
& \frac{R}{n}               & 0                 & 1             & 0                         \cr
& 0                & -1             & 0                &    \frac{R}{n}                  \cr
 & 0                & 0                 & -\frac{R}{n}              & 0                          \cr} \delta^3(x-z).
\end{equation*}
Therefore, by using the matrix $ \left(C^{(n)}_{\alpha \beta}\right)^{-1}$ and the definition of Dirac's brackets,  the non-zero  brackets among the physical fields of  the KK-modes  are  given by 
\begin{eqnarray}
\{A^{(n)}_i(x), \Pi^j_{(n)} (z)\}_{D} &=& \delta^j{_{i}}  \delta^3 (x-z), \nonumber \\ 
\{\Pi ^5 _{(n)}(x), A_i^{(n)} (z)\}_{D} &=& \frac{R}{n} \partial_i \delta^3(x-z)\nonumber \\
\{\theta ^{(n)}(x), P_{(n)} (z)\}_{D} &=& \delta^3(x-z), 
\label{eq23a}
\end{eqnarray}
It is important to remark that these results are absent in the literature. On the other hand,  in the following section, we will use the [FJ] quantization;   we will observe  that in  [FJ]  framework  the gauge  $A_5^{(n)}\approx0$  will be interpreted as a constraint of the theory. Moreover, in such  framework  the gauge $ \Pi_5^{(n)}+\frac{n}{R}A_0^{(n)} $  will not   be  invoked. 
\section{Faddeev-Jackiw quantization for the zero-mode}
In this section we will perform the [FJ] framework for the action given in (\ref{eqlaga}), and we will obtain by  a different way  the brackets  given in  (\ref{eq21a}) and  (\ref{eq23a}),   where they were obtained by using  a pure  Dirac's method. It is important to comment that the [FJ] formulation has not been applied for theories in the context of extra dimensions, thus, in this section we perform this formulation and we will show the advantages  of   the method. \\ 
For our purposes,  first we will work with the zero-mode, thus, from the  Legendre transformation (\ref{hami})   we  identify the  first order symplectic Lagrangian  for the zero-mode given by 
\begin{eqnarray}
{\mathcal{L}}^0&=&  \Pi^{i}_{(0)} \dot{A}^{(0)}_i + P _{(0)}\dot{\theta}^{(0)} - V^{0}, 
\label{eq24ac}
\end{eqnarray} 
where $V^{0}=  \Bigg(\frac{1}{2}\Pi^{i}_{(0)}\Pi_{i}^{(0)}+\frac{1}{4m^{2}}P_{(0)}P^{(0)}+\frac{1}{4}F^{(0)}_{i j}F^{i j}_{(0)}-A^{(0)}_{0}\big(\partial_{i}\Pi^{i}_{(0)}+P_{(0)}\big)-m^{2}\left(A^{(0)}_{i}+\partial_{i}\theta^{(0)}\right)^{2} \Bigg)  $. The corresponding symplectic equations of motion are given by   \cite{16}
\begin{equation}
f_{ij}^0 \dot{\xi} ^j= \frac{\partial V^0}{\partial \xi^i}, 
\end{equation}
where the symplectic matrix is defined  by
\begin{equation}
f_{ij}^0(x,y) = \frac{\delta a_j(x)}{\delta \xi^i(y)} - \frac{\delta a_i (x)}{\delta \xi^j(y)}. 
\end{equation}
Thus, from the symplectic Lagrangian (\ref{eq24ac}), we identify  the following set of symplectic variables as $\xi^i= (A_i^{(0)}, \Pi^i_{(0)}, \theta^{(0)}, P_{(0)}, A_0^{(0)}) $  and the components of the symplectic 1-forms are $a_i =\left(\Pi^i_{(0)}, 0, P_{(0)}, 0, 0 \right)$. In  this manner, by using the symplectic variables we obtain the following symplectic matrix 
\begin{eqnarray*}
\label{eq}
f_{ij}^0(x,y) =
\left(
  \begin{array}{ccccc}
    0   &   -\delta{_{j}}^i   &   0     &   0   &   0                                                                     \\
    \delta{_{j}}^i   &   0   & 0   &   0   &   0                                                                     \\
    0   & 0   &   0     &   -1   &   0                                                                     \\
    0   &   0   &   1     &   0   &   0                                                                     \\
    0   &   0   & 0     &   0   &   0                                                                    \\
      \end{array}
\right)  \delta^3(x-y), 
\end{eqnarray*}
where we can observe that this matrix is singular. In fact, in  [FJ] framework,  this means that there are constraints for the theory. We calculate the   modes of this matrix; for this theory there is a mode  and it is    given by $v^0= (0,0,0,0, \omega^{A_0^{(0)}})$, where $\omega^{A_0^{(0)}}$ is an arbitrary function. Thus, by using this mode, we obtain the following constraint 
\begin{equation}
\Omega^0 = v^0_i \frac{\delta V^0}{ \delta \xi^i} \rightarrow \partial_{i}\Pi^{i}_{(0)}+P_{(0)} = 0, 
\label{eqw}
\end{equation}
we can observe that this  constraint is the secondary constraint obtained by means Dirac's  method given in (\ref{eq36}).  We would comment that in [FJ] framework, there are not Dirac's  primary constraints as expected.  Let us  calculate  if there are present  more constraints in the context of [FJ]. In order to archive  this aim, we write in  matrix form the following system  \cite{20}
\begin{eqnarray}
 f^0 _{ij}\dot{\xi}^j &=& \frac{\delta V^0}{ \delta \xi}, \nonumber \\
\frac{\delta \Omega^0}{\delta \xi^i } \dot{\xi}^i&=&0,
\label{eq28b}
\end{eqnarray}
by using the symplectic variables and $V^0$,  that matrix has the explicit  form 
\begin{eqnarray}
\label{eq29k}
F_{ij}{(x,y)} =
\left(
  \begin{array}{ccccc}
    0   &   -\delta{_{j}}^i   &   0     &   0   &   0                                                                     \\
    \delta{_{j}}^i   &   0   & 0   &   0   &   0                                                                    \\
    0   & 0   &   0     &   -1   &   0                                                                     \\
    0   &   0   &   1     &   0   &   0                                                                     \\
    0& 0& 0 & 0 & 0 \\
    0   &   \partial_i   & 0     &   1   &   0                                                                    \\
      \end{array}
\right)  \delta^3(x-y),
\end{eqnarray}
thus, we can observe that (\ref{eq29k})  is not a squared matrix  as is expected, however, it has a  mode given by  $(v^1)_i^T= \left(-\partial_i \omega^{A_i^{(0)}}, 0, \omega^{A_i^{(0)}}, 0, 0, \omega^{A_i^{(0)}}    \right)$. This mode,  is used  in order to obtain more constraints, thus, we  calculate  the following contraction \cite{20}
\begin{equation}
(v^1)_i^T Z_i=0,
\label{eq30aa}
\end{equation}
where 
\begin{equation}
Z_i= \left(\begin{array}{c} \frac{\delta V^0}{\delta \xi^i}\\0\end{array}\right).
\label{eq31b}
\end{equation}
By performing the contraction with the  mode $(v^1)_i^T$, we find that  (\ref{eq30aa}) is an identity,  therefore, in [FJ] framework  there are not more constraints for the theory under study. \\
By  following with the method, in order to construct a new symplectic Lagrangian containing  the information of the constraint  obtained in (\ref{eqw}),   we introduce a Lagrangian multiplier associated to   the constraint  $\Omega^0$, namely $\rho^{(0)}$ , and we obtain the following symplectic  Lagrangian
\begin{eqnarray}
{\mathcal{L}}^1&=&  \Pi^{i}_{(0)} \dot{A}^{(0)}_i + P _{(0)}\dot{\theta}^{(0)} - ( \partial_{i}\Pi^{i}_{(0)}+P_{(0)}) \dot{\rho}^{(0)} - V^{1}, 
\label{eq28a}
\end{eqnarray} 
where $V^1= V{^0}|_{\Omega^0=0}= \Bigg(\frac{1}{2}\Pi^{i}_{(0)}\Pi^{i}_{(0)}+\frac{1}{4m^{2}}P_{(0)}P_{(0)}+\frac{1}{4}F^{(0)}_{i j}F^{i j}_{(0)}-m^{2}\left(A^{(0)}_{i}+\partial_{i}\theta^{(0)}\right)^{2} \Bigg)  $. Now we will consider $\rho^{(0)}$ as a symplectic variable, thus our new set of symplectic variables are given by $\xi^{1i} = (A_i^{(0)}, \Pi^i_{(0)}, \theta^{(0)}, P_{(0)}, \rho^{(0)}) $ and the new symplectic 1-forms are $a^1_i =\left(\Pi^i_{(0)}, 0, P_{(0)}, 0, - ( \partial_{i}\Pi^{i}_{(0)}+P_{(0)})\right)$. In this manner,  with these symplectic variables, we obtain the following symplectic matrix given by 
\begin{eqnarray}
\label{eq33a}
f_{ij}^1(x,y) =
\left(
  \begin{array}{ccccc}
    0   &   -\delta{_{j}}^i   &   0     &   0   &   0                                                                     \\
    \delta{_{j}}^i   &   0   & 0   &   0   &   -\partial_i                                                                    \\
    0   & 0   &   0     &   -1   &   0                                                                     \\
    0   &   0   &   1     &   0   &   -1                                                                     \\
    0   &   \partial_i   & 0     &   1   &   0                                                                    \\
      \end{array}
\right)  \delta^3(x-y).
\end{eqnarray}
We are able to observe that $f_{ij}^1(x,y) $ is singular, however, we have showed that there are not more constraints; the noninvertibility  of (\ref{eq33a})  means that the theory has a gauge symmetry. Hence, we choose the following ( gauge condition) constraint  $\Omega^2= \partial^iA_i^{(0)}=0$,  and we introduce a new Lagrange multiplier $\eta_{(0)}$  for  constructing  the following symplectic  Lagrangian 
\begin{eqnarray}
{\mathcal{L}}^2&=&  \Pi^{i}_{(0)} \dot{A}^{(0)}_i + P _{(0)}\dot{\theta}^{(0)} - ( \partial_{i}\Pi^{i}_{(0)}+P_{(0)}) \dot{\rho}^{(0)} - (\partial^iA_i^{(0)}) \dot{\eta}_{(0)}- V^{2}, 
\label{eq30a}
\end{eqnarray} 
where $V^2= V^{1}|_{\Omega^2=0}$. From the symplectic Lagrangian (\ref{eq30a}),  we take the following symplectic variables $\xi^{2i} = (A_i^{(0)}, \Pi^i_{(0)}, \theta^{(0)}, P_{(0)}, \rho^{(0)}, \eta_{(0)}) $ and the corresponding symplectic 1-forms are 
$a^2_i =\left(\Pi^i_{(0)}, 0, P_{(0)}, 0, - ( \partial_{i}\Pi^{i}_{(0)}+P_{(0)}),-\partial^iA_i^{(0)} \right)$. In this way, by using these symplectic variables, we obtain the following symplectic matrix
\begin{eqnarray}
\label{eq29a}
f_{ij}^2(x,y) =
\left(
  \begin{array}{cccccc}
    0   &   -\delta{_{j}}^i   &   0     &   0   &   0 & -\partial_i                                                                   \\
    \delta{_{j}}^i   &   0   & 0   &   0   &   -\partial_i   &0                                                                 \\
    0   & 0   &   0     &   -1   &   0                                   &0                                  \\
    0   &   0   &   1     &   0   &   -1                                  &0                                   \\
    0   &   \partial_i   & 0     &   1   &   0                          &0                                          \\
     \partial_i   &  0  & 0     &   0   &   0                          &0                                         \\
      \end{array}
\right)  \delta^3(x-y),
\end{eqnarray}
we observe that  $f_{ij}^2(x,y)$ is not singular, hence, it is an invertible matrix. After a long but straightforward calculation,   the inverse of $f_{ij}^2(x,y)$ is given by 
\begin{eqnarray}
\label{eq32a}
(f_{ij}{^2}(x,y))^{-1} =
\left(
  \begin{array}{cccccc}
    0   &   \delta{_{j}}^i - \frac{\partial^i \partial_ j}{ \nabla^2}   &   0     &   0   &   0 & \frac{\partial^i }{\nabla^2}                                                                   \\
    -\delta{_{j}}^i + \frac{\partial^i \partial_ j}{ \nabla^2}   &   0   &   \frac{\partial^i }{\nabla^2}    &   0   &    \frac{\partial^i }{\nabla^2}     &0                                                                 \\
    0   & - \frac{\partial^i }{\nabla^2}    &   0     &   1   &   0                                   & \frac{1 }{\nabla^2}                                   \\
    0   &   0   &   -1     &   0   &   0                                  &0                                   \\
    0   &   -\frac{\partial^i }{\nabla^2}   & 0     &   0   &   0                          &\frac{1}{\nabla^2}                                           \\
     -\frac{\partial^i }{\nabla^2}   &  0  & -\frac{1}{\nabla^2}     &   0   &   -\frac{1 }{\nabla^2}                          &0                                         \\
      \end{array}
\right)  \delta^3(x-y).
\end{eqnarray}
Therefore, from (\ref{eq32a}) it is possible to identify  the following [FJ] generalized brackets  given by 
\begin{eqnarray}
\{ \xi^2_i (x), \xi^2_j (y)\}_{FJ} &=&(f_{ij}{^2}(x,y))^{-1}.  
\end{eqnarray}
Thus, from (\ref{eq32a}) we obtain 
\begin{eqnarray}
\{A^{(0)}_i(x), \Pi^j _{(0)}(y)\}_{FJ} &=& \left( \delta^j{_{i}} - \frac{\partial ^j \partial_i}{ \nabla ^2} \right) \delta^3 (x-y), \nonumber \\ 
\{P(x)_{(0)}, \theta(y)^{(0)}\}_{FJ} &=& - \delta^3 (x-y), \nonumber\\
\{\Pi^i_{(0)} (x), \theta(y) ^{(0)}\}_{FJ} &=& \frac{\partial^i}{\nabla^2} \delta^3(x-y),
\label{eq34a}
\end{eqnarray}
we observe that the generalized [FJ] brackets are equivalent  with those given in (\ref{eq21a}) obtained by using a pure Dirac's method.  In addition, the quantization of the theory is done by the replacement  of classical [FJ] brackets  to commutators 
\begin{equation}
\{\xi_i^2 (x), \xi_j^2 (y) \}_{FJ} \longrightarrow  -\frac{i}{\hbar} [ \widehat{\xi}_i^2 (x),  \widehat{\xi}_j^2 (y)  ], 
\end{equation}
where  $\widehat{\xi}_i^2 (x)$  are  the quantum operators  of the   dynamical variables. Therefore, in [FJ] framework, there are the following constraints 
\begin{eqnarray}
\psi_1^{(0)}&=& \partial^i A_i^{(0)} = 0, \nonumber\\ 
\psi_2^{(0)} &=&\partial_{i}\Pi^{i}_{(0)}+P_{(0)} = 0,
 \end{eqnarray}
 there are less constraints with respect   Dirac's framework  as expected; there are not present Dirac's primary constraints. We would remark that all these results are absent in the literature. 
\section{Faddieev-Jackiw quantization for the KK-modes}
Now we will obtain the generalized  [FJ] brackets  for the KK-modes.  For our aims, from (\ref{hami}) we identify  the symplectic Lagrangian for the KK-modes  given in the following expression 
\begin{eqnarray}
{\mathcal{L}}^{0}&=&  \Pi^{i}_{(n)} \dot{A}^{(n)}_i + P _{(n)}\dot{\theta}^{(n)} + \Pi^5_{(n)} \dot{A}_5^{(n)}- V^{0}, 
\label{eq24a}
\end{eqnarray} 
where  
\begin{eqnarray}
V^0&=& \sum_{n=1}^{\infty}\Bigg[\frac{1}{2}\Pi^{i}_{(n)}\Pi^{i}_{(n)}+\frac{1}{4m^{2}}P_{(n)}P_{(n)}+\frac{1}{4}F^{(n)}_{i j}F^{i j}_{(n)}-A^{(n)}_{0}\Big(\partial_{i}\Pi^{i}_{(n)}+\frac{n}{R}\Pi^{5}_{(n)}+P_{(n)}\Big)
-m^{2}\left(A^{(n)}_{i}+\partial_{i}\theta^{(n)}\right)^{2}\nonumber\\
& &+\frac{1}{2}\Pi^{(n)}_{5}\Pi^{(n)}_{5}+\frac{1}{2}\big(\partial_{i}A^{(n)}_{5}+\frac{n}{R}A^{(n)}_{i}\big)^{2}
-m^{2}\left(A^{(n)}_{5}-\frac{n}{R}\theta^{(n)}\right)^{2}\Bigg]\Bigg).
\label{eq40c}
\end{eqnarray}
Thus, we identify the following  symplectic dynamical variables $\xi^i= \left( A_i^{(n)}, \Pi^i_{(n)}, A_5^{(n)} ,\Pi ^5_{(n)}, \theta^{(n)}, P_{(n)}, A_0^{(n)} \right)$  and the components of the symplectic 1-forms are $a_i =\left(\Pi^i_{(n)}, 0,  \Pi^5_{(n)}, 0, P_{(n)}, 0, 0 \right)$. In  this manner, we obtain the following symplectic matrix 
\begin{eqnarray*}
\label{eq37a}
f_{ij}^0(x,y) =
\left(
  \begin{array}{ccccccc}
    0   &   -\delta{_{j}}^i   &   0     &   0   &   0    &0  &0                                                                 \\
    \delta{_{j}}^i   &   0   & 0   &   0   &   0                  &0  &0                                                       \\
    0   & 0   &   0     &   -1   &   0                                      &0 &0                                \\
    0   &   0   &   1     &   0   &   0                                     &0 &0                                  \\
    0   &   0   & 0     &   0   &   0                                           &-1 &0                           \\
    0   &   0   & 0     &   0   &   1                                           &0 &0                            \\
    0   &   0   & 0     &   0   &   0                                           &0 &0                        \\
      \end{array}
\right)  \delta^3(x-y), 
\end{eqnarray*}
we observe that  this matrix is singular, and has the following  mode  $v^0= (0, 0, 0, 0, 0, 0, \omega^{A_0^{(n)}})$, where $\omega^{A_0^{(n)}}$ is an arbitrary function. Thus, by using this vector, we obtain the following constraint 
\begin{equation}
\Omega^0 _{(n)}= v^0_i \frac{\delta V^0}{ \delta \xi^i} \rightarrow \partial_{i}\Pi^{i}_{(n)}+\frac{n}{R} \Pi ^5_{(n)}+ P_{(n)} = 0,
\label{eq41c}
\end{equation}
this constraint corresponds to the secondary constraint  obtained by  Dirac's  method. Now, we will compute  if there are more constraints. In fact, by considering  (\ref{eq40c}) and (\ref{eq41c}) into  (\ref{eq28b}),   we find that  for the KK-modes  the  matrix  (\ref{eq28b}) takes the form 
\begin{eqnarray}
\label{eq42b}
F_{ij}{(x,y)} =
\left(
  \begin{array}{ccccccc}
    0   &   -\delta{_{j}}^i   &   0     &   0   &   0 &0   &0                                                                    \\
    \delta{_{j}}^i   &   0   & 0   &   0   &   0         &0   &0                                                           \\
    0   & 0   &   0    &   -1   &   0                       &0   &0                                              \\
    0   &   0   &   1     &   0   &   0                      &0  &0                                               \\
    0& 0& 0 & 0 & 0                                            &-1   &0                   \\
    0   &   0   & 0     &   0   &   1               &0   &0                                                     \\
        0   &   0   & 0     &   0   &   0               &0   &0                                                     \\
    0   &   \partial_i   & 0     &  \frac{n}{R}   &   0               &1   &0  \\
      \end{array}
\right)  \delta^3(x-y),
\end{eqnarray}
that matrix has a  mode given by $(v^1)_i^T= \left(-\partial_i \omega^{A_i^{(n)}}, 0, -\frac{n}{R} \omega^{A_i^{(n)}}, 0, -\omega^{A_i^{(n)}}, 0, 0,  -\omega^{A_i^{(n)}}    \right)$, by using this mode in (\ref{eq30aa}),  we obtain an identity. Therefore, there are not more constraints for the KK-modes. \\
With all that  information obtained at the moment,  we introduce $\rho^{(n)}$ as a Lagrangian multiplier associated for the constraint $\Omega_{(n)}^0$,  thus, we  construct a new Lagrangian given by 
\begin{eqnarray}
{\mathcal{L}}^1&=&   \Pi^{i}_{(n)} \dot{A}^{(n)}_i + P _{(0)}\dot{\theta}^{(n)} + \Pi^5_{(n)} \dot{A}_5^{(n)} - ( \partial_{i}\Pi^{i}_{(n)}+\frac{n}{R} \Pi ^5_{(n)}+ P_{(n)}) \dot{\rho}^{(n)} - V^{1}, 
\label{eq28a}
\end{eqnarray} 
where 
\begin{eqnarray*}
V^1= V{^0}|_{\Omega^0=0}&=& \sum_{n=1}^{\infty}\Bigg[\frac{1}{2}\Pi^{i}_{(n)}\Pi^{i}_{(n)}+\frac{1}{4m^{2}}P_{(n)}P_{(n)}+\frac{1}{4}F^{(n)}_{i j}F^{i j}_{(n)}  \nonumber \\
&-& m^{2}\left(A^{(n)}_{i}+\partial_{i}\theta^{(n)}\right)^{2}
 +\frac{1}{2}\Pi^{(n)}_{5}\Pi^{(n)}_{5}+\frac{1}{2}\big(\partial_{i}A^{(n)}_{5}+\frac{n}{R}A^{(n)}_{i}\big)^{2} \nonumber \\
&-&m^{2}\left(A^{(n)}_{5}-\frac{n}{R}\theta^{(n)}\right)^{2}\Bigg].
\end{eqnarray*}
Now we will consider to $\rho^{(n)}$ as a symplectic variable and  our new set of  variables are given by $\xi^i= \left( A_i^{(n)}, \Pi^i_{(n)}, A_5^{(n)} ,\Pi ^5_{(n)}, \theta^{(n)}, P_{(n)}, \rho^{(n)} \right)$  and the new symplectic 1-forms are $a_i =\left(\Pi^i_{(n)}, 0,  \Pi^5_{(n)}, 0, P_{(n)}, 0, - ( \partial_{i}\Pi^{i}_{(n)}+\frac{n}{R} \Pi ^5_{(n)}+ P_{(n)}) \right)$. In this manner,  by using these symplectic variables, we obtain the following symplectic matrix 
\begin{eqnarray*}
\label{eq39ja}
f_{ij}^1(x,y) =
\left(
  \begin{array}{ccccccc}
    0   &   -\delta{_{j}}^i   &   0     &   0   &   0    &0  &0                                                                 \\
    \delta{_{j}}^i   &   0   & 0   &   0   &   0                  &0  &-\partial_i                                                      \\
    0   & 0   &   0     &   -1   &   0                                      &0 &0                                \\
    0   &   0   &   1     &   0   &   0                                     &0 &-\frac{n}{R}                                  \\
    0   &   0   & 0     &   0   &   0                                           &-1 &0                           \\
    0   &   0   & 0     &   0   &   1                                           &0 &0                            \\
    0   &   \partial_i   & 0     &   \frac{n}{R}    &   0                                           &0 &0                        \\
      \end{array}
\right)  \delta^3(x-y), 
\end{eqnarray*}
we are able to observe that $f_{ij}^1(x,y) $ is not an invertible matrix, this means that the theory has a gauge symmetry. Hence, in order to obtain the results given in (\ref{eq23a})  we chose the following constraint condition $\Omega^{2(n)}= A_5^{(n)}=0$. The lector could ask;  why it is not  taken into account the gauge   $\Pi_5^{(n)} +\frac{n}{R} A_0^{(n)}$  as is used in  Dirac's method?. The answer is that with that gauge,  the matrix $f_{ij}^1(x,y) $ is not invertible,  because we do not expect more constraints, then the [FJ] formalism must ends;   the constraint $\Omega^{2(n)}$ makes invertible to $f_{ij}^1(x,y) $ and  this allows us   to finish with our analysis. Hence, by introducing a  new Lagrange multiplier  associated to $\Omega^{2(n)}$ we  construct the following symplectic  Lagrangian 
\begin{eqnarray}
{\mathcal{L}}^2&=&   \Pi^{i}_{(n)} \dot{A}^{(n)}_i + P _{(0)}\dot{\theta}^{(n)} + \Pi^5_{(n)} \dot{A}_5^{(n)} - ( \partial_{i}\Pi^{i}_{(n)}+\frac{n}{R} \Pi ^5_{(n)}+ P_{(n)}) \dot{\rho} ^{(n)}- A_5^{(n) }\dot{\eta}_{(n)} - V^{2}, 
\label{eq39a}
\end{eqnarray} 
where 
\begin{eqnarray*}
V^2= V{^1}|_{\Omega^{2(n)}=0}&=& \sum_{n=1}^{\infty}\Bigg[\frac{1}{2}\Pi^{i}_{(n)}\Pi^{i}_{(n)}+\frac{1}{4m^{2}}P_{(n)}P_{(n)}+\frac{1}{4}F^{(n)}_{i j}F^{i j}_{(n)}  \nonumber \\
&-& m^{2}\left(A^{(n)}_{i}+\partial_{i}\theta^{(n)}\right)^{2}
 +\frac{1}{2}\big(\frac{n}{R}A^{(n)}_{i}\big)^{2} 
-m^{2}\left(\frac{n}{R}\theta^{(n)}\right)^{2}\Bigg].
\end{eqnarray*}
Therefore, we identify the following symplectic variables   $\xi^i= \left( A_i^{(n)}, \Pi^i_{(n)}, A_5^{(n)} ,\Pi ^5_{(n)}, \theta^{(n)}, P_{(n)}, \rho^{(n)}, \eta_{(n)}  \right)$  and the corresponding 1-forms  $a_i =\left(\Pi^i_{(n)}, 0,  \Pi^5_{(n)}, 0, P_{(n)}, 0, - ( \partial_{i}\Pi^{i}_{(n)}+\frac{n}{R} \Pi ^5_{(n)}+ P_{(n)}), -A_5^{(n)} \right)$. Thus, by using these symplectic variables we obtain the following symplectic matrix 
\begin{eqnarray}
\label{eq39a}
f_{ij}^2(x,y) =
\left(
  \begin{array}{cccccccc}
    0   &   -\delta{_{j}}^i   &   0     &   0   &   0    &0  &0  &0                                                               \\
    \delta{_{j}}^i   &   0   & 0   &   0   &   0                  &0  &-\partial_i        &0                                              \\
    0   & 0   &   0     &   -1   &   0                                      &0 &0                    &-1            \\
    0   &   0   &   1     &   0   &   0                                     &0 &-\frac{n}{R}         &0                         \\
    0   &   0   & 0     &   0   &   0                                           &-1 &0                          &0 \\
    0   &   0   & 0     &   0   &   1                                           &0 &-1                            &0\\
    0   &   \partial_i   & 0     &   \frac{n}{R}    &   0                                           &1 &0   &0                     \\
    0   &   0 & 1     &  0    &   0                                           &0 &0        &0                \\
      \end{array}
\right)  \delta^3(x-y).
\end{eqnarray}
We observe that  this  matrix is not singular, therefore we can calculate its inverse. The inverse of the matrix $f_{ij}^2(x,y)$ is given by 
\begin{eqnarray}
\label{eq41a}
(f_{ij}^2(x,y) ) ^{-1} =
\left(
  \begin{array}{cccccccc}
    0   &   \delta{_{j}}^i   &   0     &   -\frac{R}{n} \partial^i    &   0    &0  &0  &  \frac{R}{n}  \partial^i                                                            \\
   - \delta{_{j}}^i   &   0   & 0   &   0   &   0                  &0  &0        &0                                              \\
    0   & 0   &   0     &   0   &   0                                      &0 &0                    &1            \\
    \frac{R}{n} \partial^i    &   0   &   0     &   0   &   \frac{R}{n}       &0 &\frac{R}{n}         &0                         \\
    0   &   0   & 0     &  - \frac{R}{n}    &   0                                           &1 &0                          &\frac{R}{n}   \\
    0   &   0   & 0     &   0   &  - 1                                           &0 &0                            &0\\
    0   &   0  & 0     &   -\frac{R}{n}    &   0                                           &0 &0   &\frac{R}{n}                      \\
    -\frac{R}{n}  \partial^i    &   0 & -1     &  0    &   -\frac{R}{n}                                             &0 &-\frac{R}{n}         &0                \\
      \end{array}
\right)  \delta^3(x-y), 
\end{eqnarray}
In this manner, from (\ref{eq41a}) we obtain the following generalized [FJ] brackets among the physical fields  
\begin{eqnarray}
\{A^{(n)}_i(x), \Pi^j_{(n)} (y)\}_{FJ} &=& \delta^j{_{i}}  \delta^3 (x-y), \nonumber \\ 
\{\Pi ^5 _{(n)}(x), A_i^{(n)} (y)\}_{FJ} &=& \frac{R}{n} \partial_i \delta^3(x-y) , \nonumber \\
\{\theta ^{(n)}(x), P_{(n)} (y)\}_{FJ} &=&\delta^3(x-y),
\label{eq42a}
\end{eqnarray}
where we can observe that  these   brackets are the same with those  obtained in (\ref{eq23a}) by means of Dirac's framework.  It is important to comment that in Dirac's procedure, $A_5^{(n)}$ are  identified as  pseudo-Goldston bosons; in [FJ] formalism  the fields $ A_5^{(n)}$ are  identified as   constraints for the theory, this gauge allowed us end our [FJ] analysis. Hence, in [FJ] method we obtain for the KK-modes the following constraints 
\begin{eqnarray}
\partial_{i}\Pi^{i}_{(n)}+\frac{n}{R} \Pi ^5_{(n)}+ P_{(n)}& =& 0, \nonumber \\
A^{(n)}_5&=&0.
\end{eqnarray}

Therefore, we have obtained the quantum brackets   for the zero and for  KK-modes  by means  two  different approaches.  
\\
\section{Conclussions and Prospects}
In this paper,  the Hamiltonian and the [FJ] analysis for a five-dimensional St{\"{u}}eckelberg's  theory in the context of extra dimensions has been performed. Respect to  the Hamiltonian formalism,   we obtained the complete canonical description of the theory. After performing  the  compactification of the fifth dimension on a $S^1/\mathbf{Z_2}$ orbifold, we found that  the  theory is composed by  a four-dimensional  St{\"{u}}eckelberg  theory identified with the  zero-mode plus  a tower of  KK-modes. We report the complete constraints program,  we found that  the theory  has only  first class constraints. From the gauge transformations of the theory and fixing  a particular gauge for the gauge parameters,  we identified massive vector fields,   massive scalar fields and   the fields  $A_5^{(n)}$ are identified as   pseudo-Goldston bosons just as is present in five-dimensional  Maxwell theory \cite{17}. Furthermore,  we   constructed  the fundamental  Dirac's   brackets of the  zero modes  and the Dirac's brackets for the  KK-modes.\\
On the other hand, we performed the [FJ] quantization for the theory under study. We calculate the generalized [FJ] brackets  and we showed that both  [FJ] brackets and Dirac's brackets are the same. We found  in the context of [FJ] that  the fields   $A_5^{(n)}$ are identified as    constraints  of  the theory,  and they are not absorbed as it is present in Dirac's method being identified as pseudo-Goldston bosons. Furthermore,  we could observe that we arrived to the constraints and  the generalized brackets in less steps than Dirac's method, this means that [FJ] is in particular,  for the theory  under study,  more economic than Dirac's procedure \cite{21a}. In this manner,  we have stablished  all the elements for studying the quantization aspects, for example,  we can use Dirac's brackets or  [FJ] generalized brackets  for studying  the observables of the theory that  could be amenable to test. On the other hand,  by using those brackets we also could  calculate the propagators  among gauge  fields and carryout the quantization by means of  canonical approach or by means  [FJ] framework, however, all these ideas are in progress and will be the subject of forthcoming works.       \\ 
\noindent \textbf{Acknowledgements}\\[1ex]
This work was supported by CONACyT under Grant No. CB-2010/157641. We would like to thank R. Cartas-Fuentevilla for discussion on the subject and reading the manuscript. \\

\end{document}